\documentclass[twocolumn,superscriptaddress,preprintnumbers,amsmath,amssymb,longbibliography]{revtex4-1}
\usepackage[format=plain, justification=justified,singlelinecheck=false,labelfont=bf]{caption}
\usepackage{amsmath}
\usepackage{amssymb}
\usepackage{dcolumn}
\usepackage{graphicx}
\usepackage{subfigure}
\usepackage[american]{circuitikz}
\usepackage{tikz}
\usepackage{bm}
\usepackage{epstopdf}
\usepackage{threeparttable}
\usepackage{float}
\usepackage{color}
\usepackage{mathtools} 
\usepackage{mathtools}
\usepackage{balance}

\begin{document}
\title{Free Energy Cost to Assemble Superlattices of Polymer-Grafted Nanoparticles}
\author{Dingning Li} 
\affiliation{Division of Natural and Applied Sciences, Duke Kunshan University, Kunshan, Jiangsu, 215300, China}
\author{Kai Zhang} 
\email{kai.zhang@dukekunshan.edu.cn}
\affiliation{Division of Natural and Applied Sciences, Duke Kunshan University, Kunshan, Jiangsu, 215300, China}
\affiliation{Data Science Research Center (DSRC), Duke Kunshan University, Kunshan, Jiangsu, 215300, China}

\date{\today}

\begin{abstract}
 Mesoparticles consisting of a hard core and a soft corona like polymer-grafted nanoparticles (PGNPs) can assemble into various superlattice structures, in which each mesoparticle assumes the shape of  the corresponding Wigner-Seitz (or Voronoi) cell. Conventional wisdom often perceives the stability of these superlattices in  a mean-field view of surface area minimization or corona entropy maximization, which lacks a molecular interpretation. We develop a simulation method to calculate the free energy cost to deform spherical PGNPs into Wigner-Seitz  polyhedra, which are then relaxed  in a certain crystalline superlattice. With this method, we successfully quantify the free energy differences between model BCC, FCC and  A15 systems of PGNPs and identify BCC as the most stable structure in most cases. Analysis of  polymer  configurations in the corona, whose boundary is blurred by chain interpenetration,  shows that the radial distribution of grafted chains and the corresponding entropy is almost identical among BCC and FCC, suggesting that the higher stability of BCC structure cannot be explained by a mean-field description of corona shape.
\end{abstract}
 
\maketitle

\section{Introduction}
Many complex properties of soft matter, such as ordered or disordered mesoscopic structures, weak interactions comparable to thermal agitation and strong response to weak perturbation, are rooted in the fact that the basic building blocks of the system are soft deformable mesoparticles~\cite{brochard2020}. 
Such soft mesoparticles, as diverse as polymer-grafted nanoparticles (PGNPs)~\cite{akcora2009},  spherical micelles of diblock
copolymers~\cite{thomas1987}, giant supramolecules~\cite{ungar2003}, ligand-capped nanocrystals~\cite{goodfellow2015}, and DNA-functionalized nanoparticles~\cite{nykypanchuk2008},  are all made of a relatively hard core  and a soft corona of chain molecules.  The chemical complexity in above examples provides us with plethora of possibilities to manipulate their self-assembly behavior and enables the fabrication of  various superlattice structures including but not limited to body-centered cubic (BCC),  face-centered cubic (FCC) and hexagonal-closed packed~\cite{si2018,girard2019}. A quantitative account of the stability of competing superlattices is of both theoretical and experimental interest, which is the prerequisite to predict equilibrium phase properties~\cite{boles2016}.
 
Although packing has been successfully applied to describe hard particles~\cite{cersonsky2018}, it does not capture the effects of soft corona. Attempts have been made to model soft  mesoparticles with simple effective potentials such as hard-core square shoulder (HCSS) potential~\cite{ziherl2000,ziherl2001,pattabhiraman2017} and inverse power law potential~\cite{travesset2015}.  Theoretical analysis based on  simple mean-field picture identifies two contributions, lattice vibration and surface minimization, to the overall crystal stability~\cite{ziherl2000,ziherl2001}.
However, interactions in these models are spherically symmetric and pair-wise additive, thus miss the intrinsic anisotropy and manybody effects of core-corona mesoparticles. In fact, one key feature of these deformable mesoparticles is that they adopt the shape of Wigner-Seitz cells (or Voronoi cells) in the superlattice~\cite{goodfellow2015,bilchak2017,reddy2018}. These polyhedral unit cells of soft mesoparticles are analogously observed in metals~\cite{lifshitz2014} and metallic nanocrystals~\cite{li2020}, but due to different physical origins.

To address the polyhedral shapes of soft mesoparticles, another important perspective was introduced, which can date back to the famous Kelvin problem on dividing space into equal partitions with minimum amount of materials~\cite{williams1968}. Regarding the Wigner-Seitz  polyhedron of BCC (truncated octahedron) and FCC (rhombic dodecahedron), the surface area  $A_{\rm BCC}$  is less than $A_{\rm FCC}$  by $0.58\%$ for a given volume. Kelvin's original solution was a modified truncated octahedron where some faces are changed to be curved. It was later recognized that, by allowing two types of polyhedra, the minimal area solution to the Kelvin's  problem becomes the Weaire-Phelan foam, which contains six  tetrakaidecahedra (Z14) and two irregular dodecahedra (Z12) in each unit cell~\cite{weaire1994,kusner1996}. This structure actually occurs in nature as in A15 clathrate~\cite{rivier1994}, which belongs to the large group of  Frank-Kasper phases~\cite{frank1959,montis2021}. The average surface area  $\bar{A}_{\rm A15}$  of the eight polyhedra in A15 unit cell is less than  $A_{\rm BCC}$ by $0.33\%$~\cite{kusner1996}.

The minimal surface consideration has found its  applicaiton in diblock copolymers, which organize into soft micelles to form  various superlattice structures~\cite{grason2003}. This soft micelle differs from other core-corona mesoparticles by its soft core made of one type of polymer segements. At different conditions, the shape of the soft core may change from spherical to polyhedral~\cite{reddy2018}.  The stability  of diblock copolymer phases is  then a tradeoff between the interfacial energy (between two blocks) and the stretching energy of chains~\cite{grason2003}.  Although BCC spherical phase was first discovered in diblock copolymers, unexpected  Frank-Kasper phases are also observed~\cite{lee2010,lee2014,kim2017,schulze2017,watanabe2020}.
Self-consistent field theory (SCFT)  predicts that conformation asymmetry and copolymer architecture (branched or not) can be tuned to favor the formation of complicated  Frank-Kasper phases~\cite{liu2016,li2017,bates2019}. However, a molecular-level  understanding of the competition between different superlattice structures is still missing.

To resolve the stability of superlattices formed by deformable mesoparticles, we propose a free energy calculation scheme based on thermodynamic integration~\cite{frenkel:2002} and apply it on polymer-grafted nanoparticles (PGNPs). Similar free energy calculations have been performed  on hydrocarbon-capped nanocrystals using thermodynamic perturbation~\cite{kaushik2013} and integration over pressure~\cite{zha2018}.
In our method, the free energy cost $F$ to assemble isolated spherical PGNPs into BCC, FCC and A15 superlattices are calculated in two steps. In step I, PGNPs are compressed into polyhedral shapes by confining walls. In step II, compressed PGNPs are relaxed on a superlattice while confining walls are removed. By implementing this methodology with  coarse-grained molecular dynamics (MD) simulation, we   quantify the stability of different superlattices in the limit of low (super)lattice vibration. We find that BCC structure is marginally more stable than FCC by $\sim1\%$ for most volume studied.  A test on one A15 structure shows it is more stable than FCC but still less stable than BCC, suggesting that minimal surface area does not always guarantee lower free energy. Close examination of polymer distribution around the nanoparticle core reveals that chain interpenetration from two PGNPs blurs the boundary of Wigner-Seitz polyhedra~\cite{midya2020}. The corresponding  entropy $S$ of corona chain  distribution is almost indistinguishable among FCC and BCC~\cite{goodfellow2015}, making the determination of superlattice stability by   $S$  alone impossible.

\section{Methods}
 \subsection{Thermodynamic Integration}
Thermodynamic integration is a standard free energy calculation method which connects   the reference state ``0'' with  the state of interest ``1'' by a reversible path denoted by a parameter, e.g. $\lambda \in [0,1]$~\cite{frenkel:2002}. If the potential energy at the two endpoints are $U_0$ and $U_1$,  then the potential energy for the coupled system at intermediate $\lambda$ can be defined as  $U(\lambda)=(1-\lambda)U_0+\lambda U_1$ and the free energy of interest is calculated by the integral along the path
\begin{equation}
	F_1 = F_0 + \int_{0}^{1}\left\langle\frac{\partial U(\lambda)}{\partial \lambda}\right\rangle_{\lambda} d\lambda
\end{equation}
where  $\left\langle \cdots \right\rangle_{\lambda} $ means to take  ensemble average in the system interacting with potential energy $U(\lambda)$.

We choose isolated PGNPs as our reference state ``0'' and  consider  a reversible thermodynamic path to assemble crystalline superlattices of PGNPs in two steps (Fig.~\ref{fig:TI}). In step I, each spherical PGNP in  isolated state is slowly compressed  into a  polyhedron (usually with  a shape of the Wigner-Seitz cell) by gradually increasing the repulsion  strength of confining walls. In step II, confined PGNPs are stacked to form the desired superlattice,  whose interactions  are slowly turned on while wall confinement is being removed. In this work, we do not calculate the reference free energy $F_0$ for the hyperthetical crystal formed by  uncompressed spherical PGNPs. We also neglect the contribution from collective vibrations of PGNPs on the superlattice, which should be small for highly compressed states. Therefore, the free energy $F$  reported in this work is the cost (per PGNP)  to assemble the superlattice  from uncompressed spherical PGNPs, neglecting lattice vibration. We break $F$ into two terms corresponding to the two steps (to be explained below)
\begin{equation}
F = F_V + F_S.
\end{equation}
\begin{center}
\includegraphics[width=0.45\textwidth]{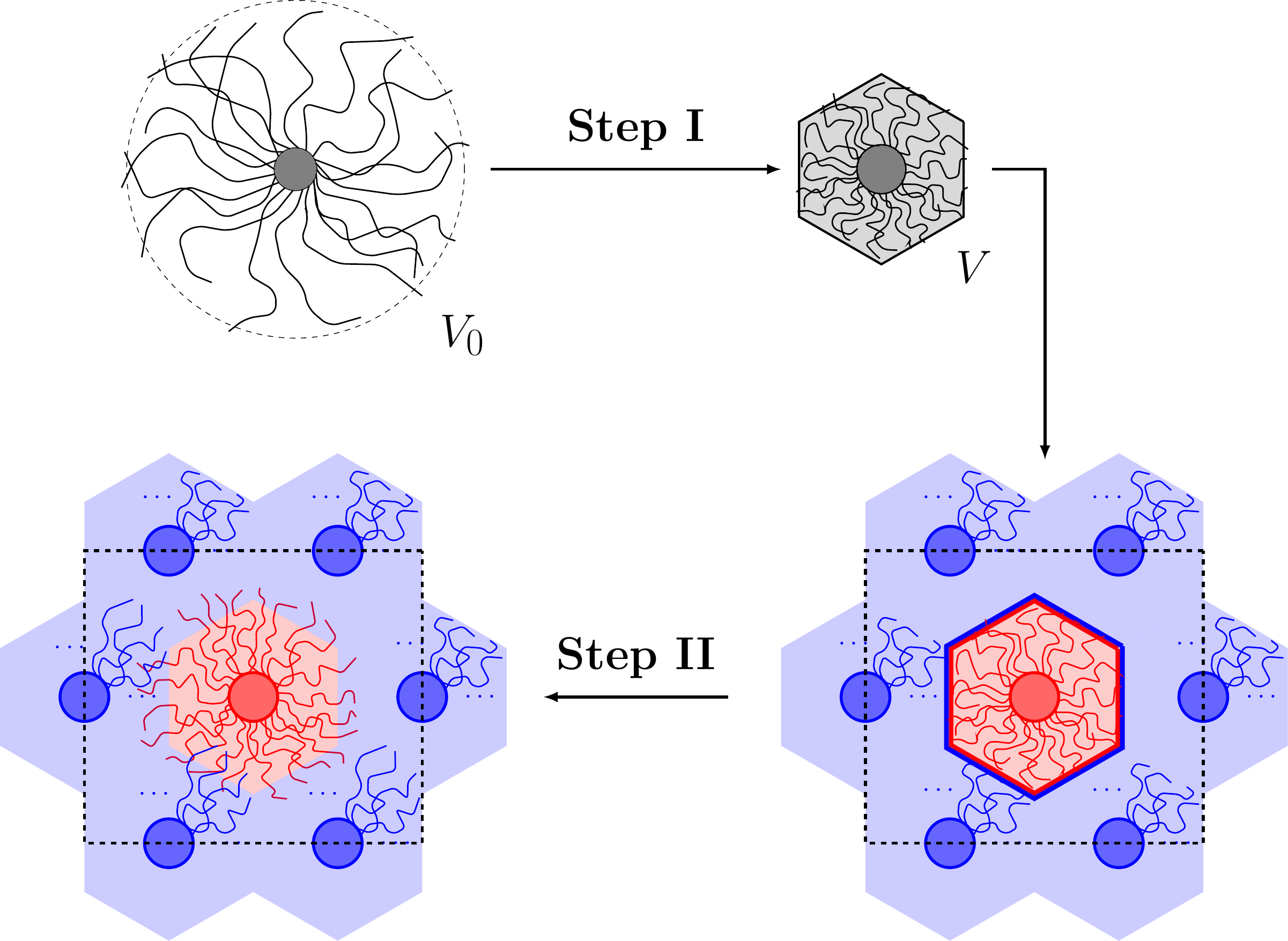}
\captionof{figure}{Schematic representation of the thermodynamic integration path to assemble crystalline superlattices  from isolated  PGNPs in two steps. Repulsive confining walls are represented by solid lines around the (central) polygon. A possible choice of the simulation box in step II under periodic boundary conditions  is shown by the dashed rectangle.}
\label{fig:TI}
\end{center}

In step I, the starting point is the system of one isolated spherical PGNP with volume $V_0$ and potential energy $U_0 = U_{\rm bond} + U_{\rm nonbond}$, where $U_{\rm bond}$ is the energy of covalent bonds in polymer chains and $U_{\rm nonbond}$  is pairwise-additive non-bonded interactions between monomers or nanoparticles. During compression along the path, the PGNP experices repulsion $W$ from several confining walls adopting the corresponding geometry of the desired superlattice.  The positions of these walls define the final volume $V<V_0$ of the PGNP after  compression. The coupled potential energy at integration parameter $\lambda$ is defined as
\begin{equation}\label{U1}
U_{\rm I}(\lambda) = U_{\rm bond} + U_{\rm nonbond} + \lambda W
\end{equation}
with $\lambda$ changing from $0$ to $1$. The free energy cost  $F_V$ in this step can be roughly associated with volume compression of the PGNP and is calculated  by the integral
\begin{equation}\label{eq:Fv}
	\begin{aligned}
F_V=\int_{0}^{1}\left\langle W\right\rangle_{\lambda}\,d\lambda.
	\end{aligned}
\end{equation}

In step II, the fully compressed PGNPs from step I are packed into the desired superlattice. Non-bonded interactions in the system can be decomposed into three parts, $U_{\rm nonbond} = U_N + U_{N'} + U_{NN'}$, where $U_N $ is the interaction energy among the $N$ particles of the central PGNP,  $U_{N'} $ is the interaction energy among the $N'$ particles of the surrounding PGNPs (usually just first nearest neighbors), and $U_{NN'}$ is  interaction energy between particles of the central PGNP and particles of the surrounding PGNPs. In addition to the walls confining the central PGNP (same as in step I), an equal number of extra walls facing outward are needed to confine the surrounding PGNPs. There is no wall between two surrounding PGNPs.  If the interaction with all the walls is $W$, then the coupled potential energy is written as
\begin{equation}\label{U2}
U_{\rm II}(\lambda) = U_{\rm bond} +  U_N + U_{N'} +  \lambda U_{NN'} + (1 - \lambda) W
\end{equation}
with $\lambda$ changing from $0$ to $1$. In practice, one might need to start from over-compressed PGNPs with smaller volume $V'<V$, which leave small gaps between them to avoid initial  overlap of polymer chains. After relaxation, these gaps are filled by polymer chains. The contribution of the free energy $F_S (<0)$ in step II to the total free energy cost per PGNP $F$ is
\begin{equation}\label{eq:Fs}
	\begin{aligned}
F_S= \frac{1}{2}\int_{0}^{1}\left\langle U_{NN'} - W\right\rangle_{\lambda}\,d\lambda,
	\end{aligned}
\end{equation}
which is related to the  relaxation of surfaces between PGNPs after walls are removed. The factor $\frac{1}{2}$ is due to the sharing of each surface by two PGNPs.

\subsection{Models and Molecular Dynamics Simulation}
We use the coarse-grained model  of polymer chains and nanoparticles, in which non-bonded interactions between a pair $(i,j)$ of monomers ($m$) and/or nanoparticles ($n$) at a distance $r_{ij}$ apart follow the repulsive  shifted Lennard-Jones (LJ) potential, when   $ r_{ij}<r_c+\Delta_{ij}$,
$u_{\rm LJ}(r_{ij})=4\epsilon\left[\left(\frac{\sigma}{r_{ij}-\Delta_{ij}}\right)^{12}-\left(\frac{\sigma}{r_{ij}-\Delta_{ij}}\right)^6\right],$
where  $\epsilon$ is the unit of energy and the monomer size $\sigma$ is the unit of length~\cite{bilchak2017}.  The choice of $\Delta_{nn} = 9\sigma$, $\Delta_{mn} = 4.5\sigma$ and $\Delta_{mm} = 0$ makes the effective size of bare nanoparticles  roughly $10\sigma$. The cutoff distance is set to $r_c = 1.12\sigma$ so that all pair interactions are purely repulsive.
Polymers are first modeled as bead-spring chains connected by finite extensible nonlinear elastic (FENE) bonds, $u_{\rm FENE}(r) = -0.5 KR_0^2 \ln [1 - (r/R_0)^2]$ $(r<R_0)$, with force constant $K=30 \epsilon/\sigma^2$ and bond length $R_0 = 1.5\sigma$~\cite{kremer1990}. We also consider semiflexible chains with the harmonic  bond angle potential $u_\theta(\theta) = K_\theta (\theta - \theta_0)^2$ added on top of FENE bonds, where $K_\theta = 300 \epsilon\cdot$rad$^{-2}$ and $\theta_0 = 109.5^\circ$.  We graft $N_c = 200$ polymers by anchoring the first monomer of each chain  on the surface of the nanoparticle with a rigid bond. The corresponding surface graft density is $\Sigma = 0.637  \sigma^{-2}$. Several chain lengths are studied with the  number of monomers per chain $N_p=20,40,60$.
\begin{center}
	\includegraphics[width=0.45\textwidth]{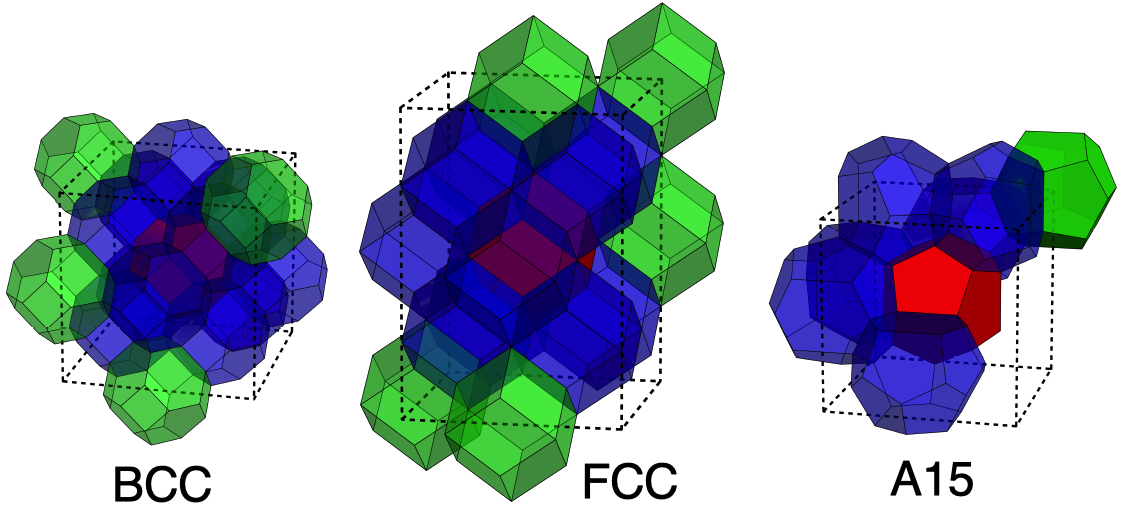}
	\captionof{figure}{Stacking of  Wigner-Seitz polyhedra  in thermodynamic integral step II to study BCC, FCC and A15 phases, respectively. Red, blue and green colors indicate central polyhedron, its first   and second nearest neighbors. For clarity, not all polyhedra in the simulation box are shown. Note that A15 has two different shapes of polyhedra:  tetrakaidecahedron Z14 (blue) and irregular dodecahedron Z12 (red and green). }
	\label{fig:box}
\end{center}

When PGNPs need to be confined  into the desired polyhedral shape, polymers  also interact with several repulsive walls around the PGNP through a harmonic potential $u_{w}(r) = 80\epsilon(r/\sigma)^2$, where $r$ is the penetration depth of a monomer into the wall surface. In a system with $Z$ walls and $N$ monomers,  the total wall energy is  $	W = \sum\limits_{i=1}^N \sum\limits_{j=1}^Z u_w(r_{ij})$. For the BCC, FCC and A15 structures under study, walls are placed at the faces of the corresponding Wigner-Seitz cells, which are  truncated octahedron (BCC), rhombic dodecahedron (FCC), tetrakaidecahedron and irregular dodecahedron (A15), respectively. Details of wall positions are described in Supporting Information.
In step I, only one PGNP  is needed under open boundary conditions. In step II, several surrounding PGNPs (not only  nearest neighbors, but also some next nearest neighbors) are needed to enclose the central one.  By properly choosing the simulation box under periodic boundary condisions, we use 16, 16 and 64 PGNPs to study BCC,  FCC and A15, respectively (Fig~\ref{fig:box}).

The LAMMPS  package is used to run constant volume molecular dynamics simulations at fixed temperature $T = 1.0 \epsilon/k_B$ with a time step $\Delta t = 0.005 \sigma \sqrt{m/\epsilon}$, where the mass $m$ is set the same for all particles~\cite{plimpton1995}. Each system is equilibrated for $\sim 10^6$ time steps before ensemble average is taken.  To  evaluate each thermodynamic integral numerically,  $n_\lambda  = 50$-$60$ $\lambda$ points are used within Gauss-Lobatto quadrature~\cite{press1992}. Two independent runs are performed for each system, whose numerical discrepancy is an order of magnitude smaller than the free energy difference between different phases studied here. All  free energy results are reported in unit of $\epsilon$.

\section{Results and discussion}
\subsection{Stability of crystalline superlattices}
We calculate the free energy cost $F$ to assemble BCC, FCC and A15 superlattices at various Wigner-Seitz (Voronoi) cell volumes $V$ for PGNPs with  flexible (FENE bonds) or semiflexible (FENE bonds plus bond angles) grafts.   The volume $V$ of the polyhedron in superlattice is compared with the original volume $V_0 = \frac{4\pi}{3} R_0^3$ of an isolated spherical PGNP, whose radius $R_0$ is estimated as the average of the location of the last monomer on each chain. The typical thermodynamic integration curves in step I and II are shown in Fig.~\ref{fig:TIcurve}. The curve drops rapidly, by orders of magnitude, close to the state without confining walls ($\lambda = 0$ in step I and $\lambda = 1$ in step II). To capture this fast change, we use $n_\lambda = 50-60$ $\lambda$ points in each integration. The calculated volume $F_V$ and surface $F_S$ free energy cost is reported as per PGNP value.  
\begin{center}
	\includegraphics[width=0.23\textwidth]{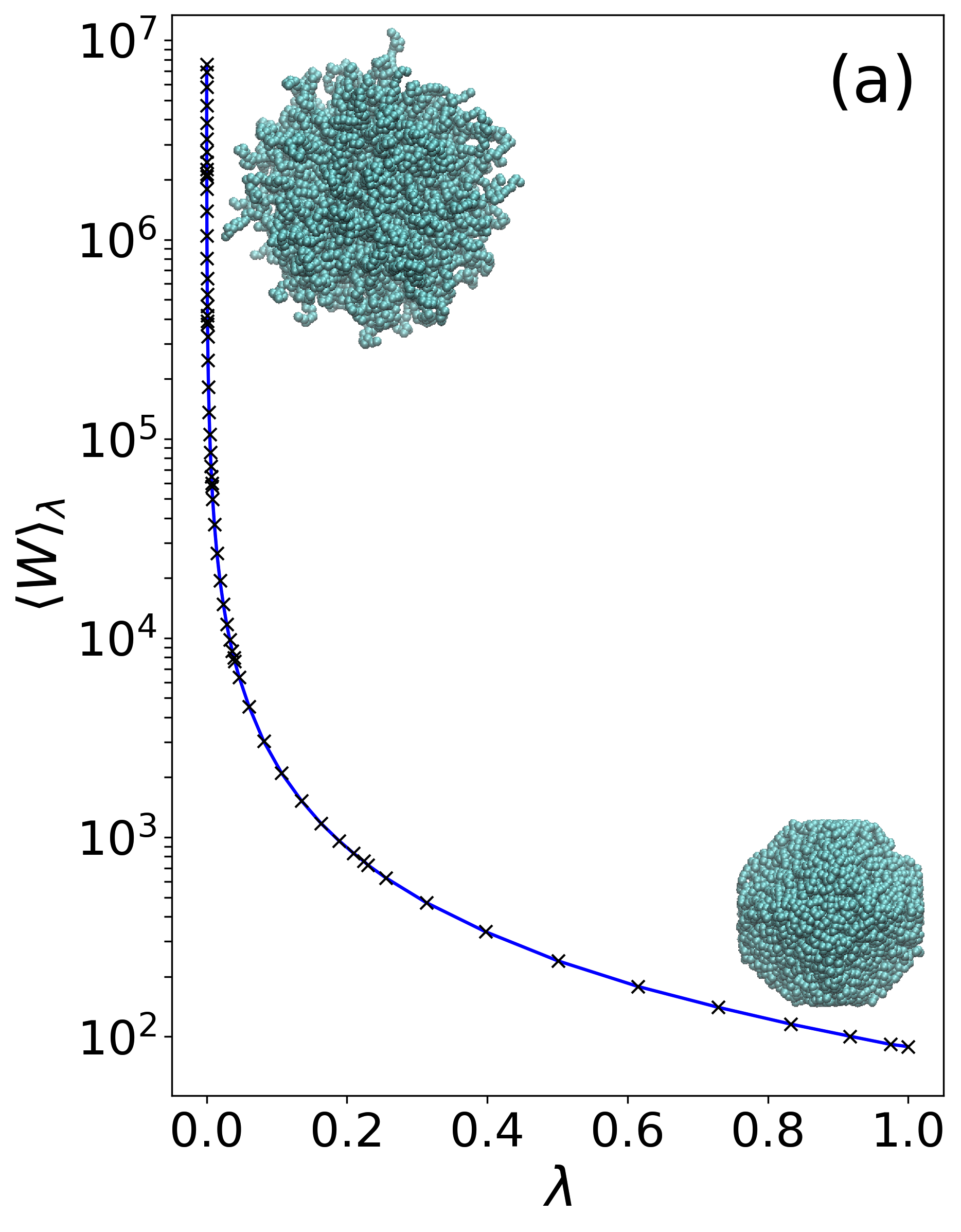}
       \includegraphics[width=0.24\textwidth]{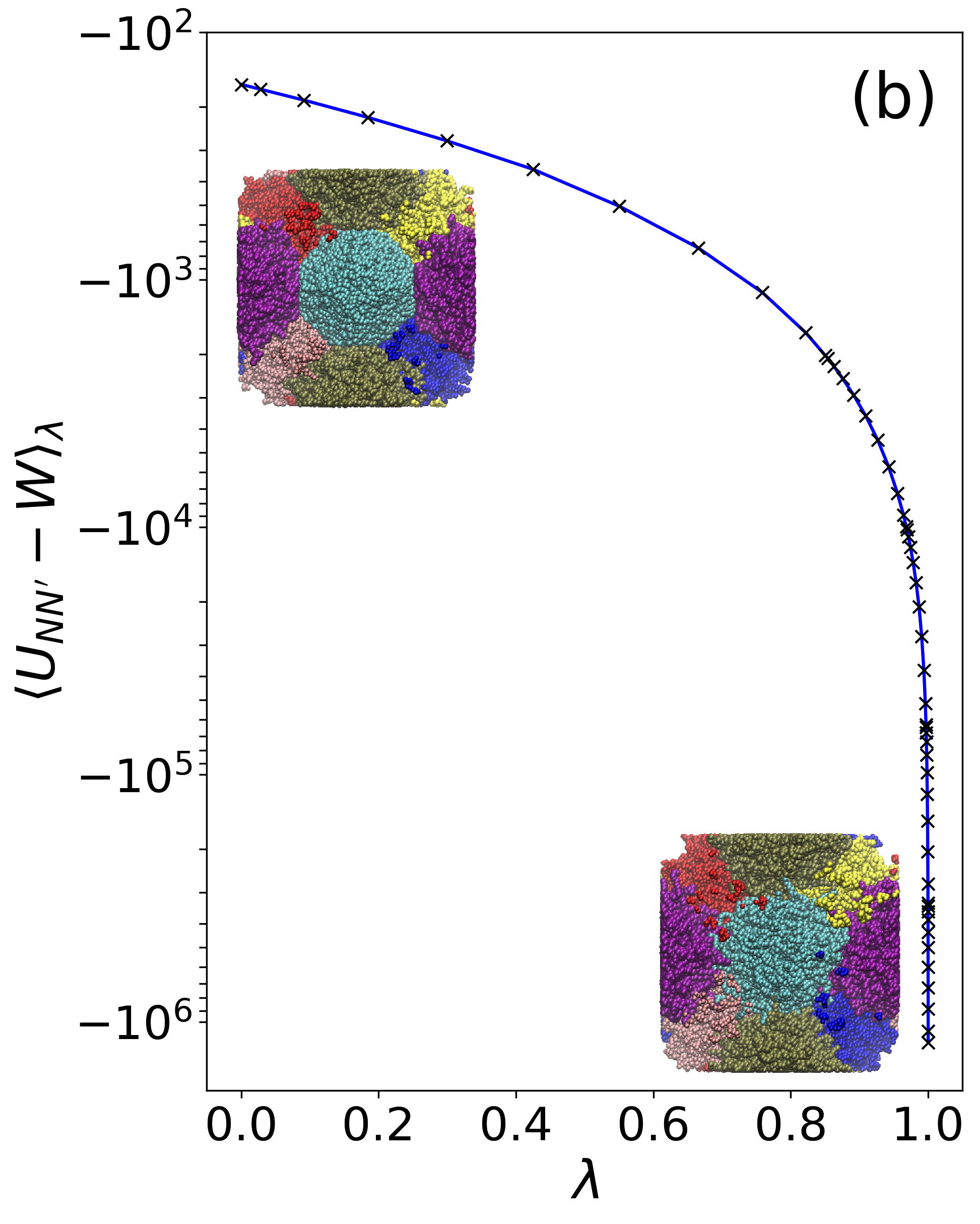}
	\captionof{figure}{Thermodynamic integration curve  with $n_\lambda=60$  data points in   step I (a) and II (b) for a BCC superlattice at $V = 18000 $ with $N_p = 40$. System configurations at the starting ($\lambda=0$) and end  ($\lambda = 1$) point of each integration are shown in   insets.  }
	\label{fig:TIcurve}
\end{center}

\begin{figure*}
\begin{center}
	\includegraphics[width=0.32\textwidth]{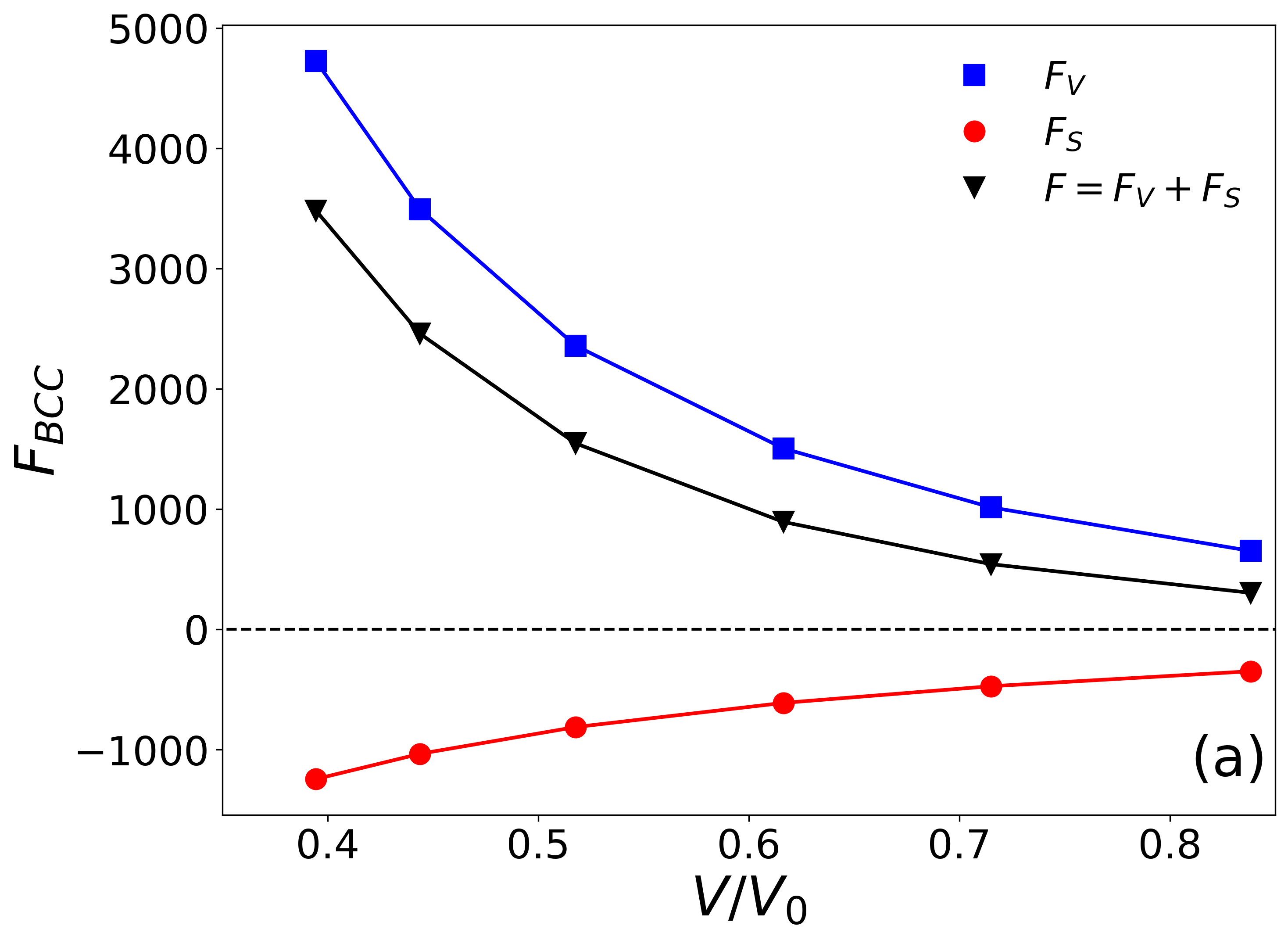}
     \includegraphics[width=0.32\textwidth]{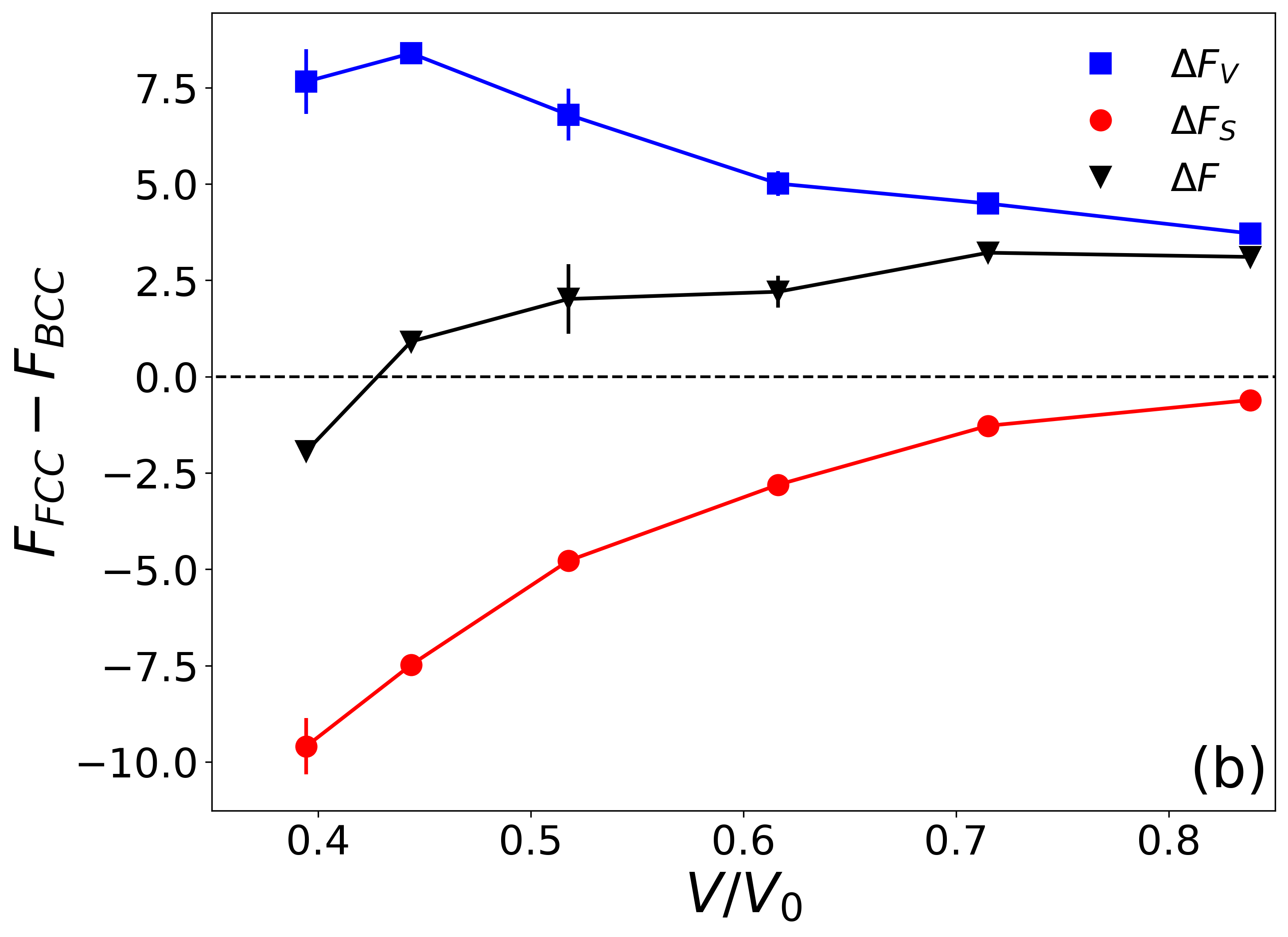}
          \includegraphics[width=0.32\textwidth]{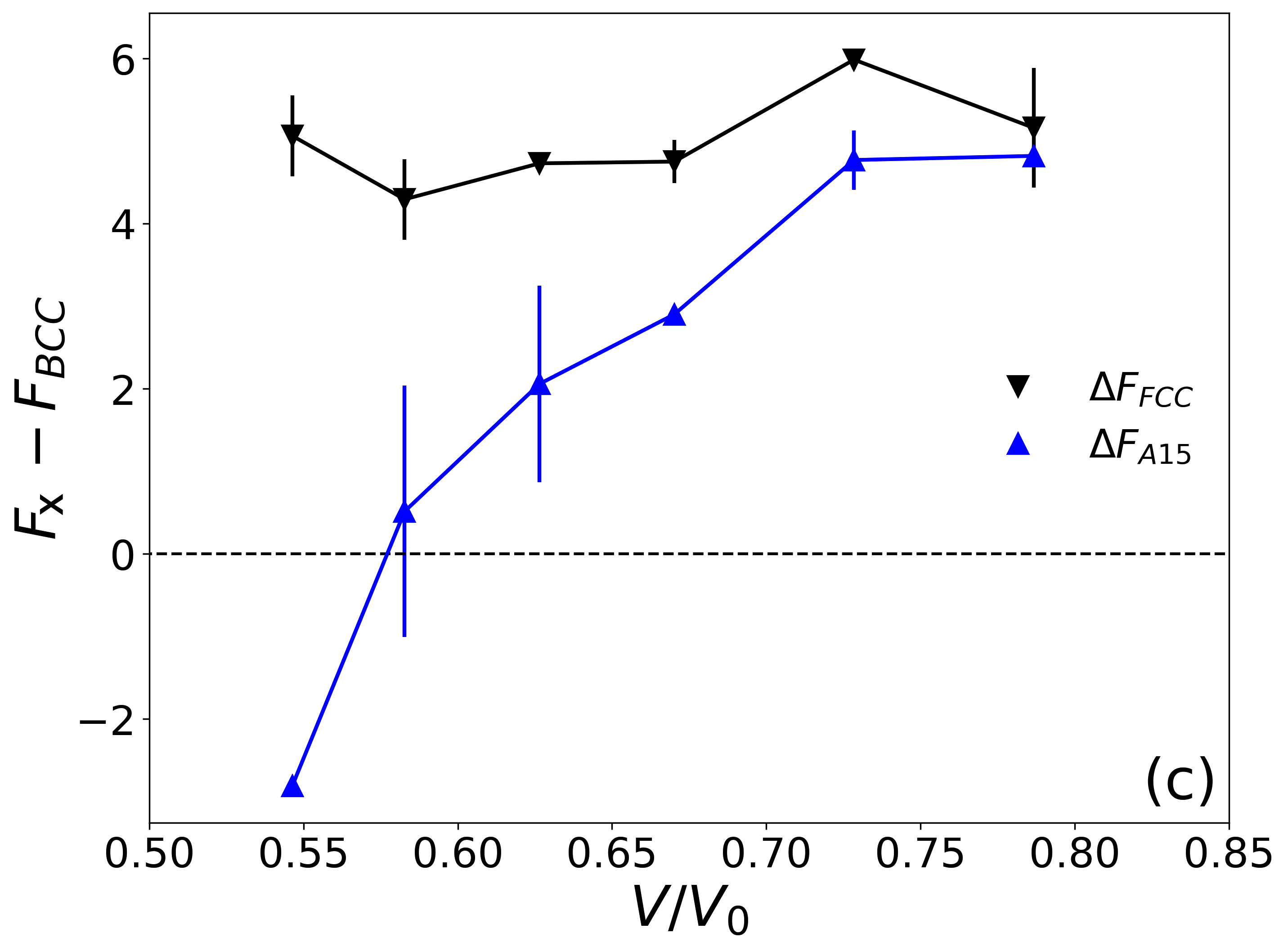}
	\captionof{figure}{(a) The volume ($F_V$), surface ($F_S$) and total ($F$) free energy cost to assemble a BCC superlattice as a function of the polyhedron   volume $V$ for $N_p = 40$ ($V_0 = 40556.9$). (b) The difference between the volume, surface and total free energy cost of a FCC superlattice  and those of the BCC superlattice  in (a) at the same $V$. (c)  The difference between the total free energy cost of a FCC  or a A15 superlattice and that of a BCC superlattice for $N_p = 20$. Polymers are flexible chains with FENE bonds. }
	\label{fig:F}
\end{center}
\end{figure*}

As expected, the volume cost $F_V$ increases as $V$ decreases because more work is needed to compress a spherical PGNP into a smaller polyhedron (Fig.~\ref{fig:F}a). The surface term $F_S$ is negative because it corresponds to the free energy gain (entropy increase) after the confined flat interface is relaxed, allowing interpenetration of polymers from two neighboring PGNPs. The overall free energy cost $F = F_V + F_S$ still increases monotonically  with decreasing $V$.

For all the chain lengths $N_p = 20, 40, 60$ studied, being flexible or semiflexible, we find  that BCC is more stable than FCC structure at large $V$ (Fig.~\ref{fig:F}b). We can first  identify a higher free energy cost $F_V$ for confined FCC as shown in Fig.~\ref{fig:F}b, but the difference   $\Delta F_V =  F_V({\rm FCC}) - F_V({\rm BCC})$ is only $\lesssim 1\%$ of $F_V({\rm BCC})$. To compare the total free energy cost $F$,  we also need to include the free energy gain $F_S (<0)$ from surface relaxation. Since  rhombic dodecahedron  has $ 0.58\%$ more surface area than  truncated octahedron, $|F_S|$ is  slightly larger in FCC than BCC. But this surface relaxation is not strong enough to compensate the volume cost  $F_V$. Combining two effects,  the free energy cost $F$ of BCC is lower than that of FCC structure by at most  $\sim1\%$ (see data in Supporting Information).  This is in agreement with previous calculations on ligand-capped nanocrystals~\cite{zha2018},  but  in contrast to the theoretical analysis of entropy penalty for chain compression/extension, which estimates that BCC can be $15\%$ more stable than FCC in ligand-capped nanocrystals~\cite{goodfellow2015}.
In the small $V$ limit, the numerical results of $F$ of FCC is close to or even smaller than that of BCC. But the relative difference, $ \frac{ F_{\rm FCC} - F_{\rm BCC}  }{  F_{\rm BCC}}$, is on the order of $0.1\%$, which is near the numerical precison of our thermodynamic integration scheme. Therefore,  there is no clear evidence to claim that FCC is more stable than BCC at small volume, although it is possible.

We also examine the stability of A15  systems with short ($N_p = 20$) flexible chains. Because the unit cell of A15 contains eight particles and  64 PGNPs are used  in step II of thermodynamic integration,  it is computationally too expensive to study longer chains under current protocol. In Voronoi tessellation of  A15 structures, the two types of Frank-Kasper polyhedra, tetrakaidecahedron (Z14) and irregular dodecahedron (Z12),  can  have different volumes in principle. Previous study on block copolymers found that the free energy of A15 is optimized, when the volumes of Z14 and Z12  are slightly different~\cite{reddy2018}. Here we focus on  the equal-cell partition as in  Weaire-Phelan foam~\cite{weaire1994}, where  the same $V$ for the  Z14 and Z12 is assumed.  The calculated the free energy $F$  shows that, although being slightly more stable than FCC, A15 is still less stable than BCC structure except at small  $V$ (Fig.~\ref{fig:F}c). Since the surface area $A$ of the Wigner-Seitz polyhedron for a given volume $V$ obeys $A_{\rm FCC} > A_{\rm BCC} > \bar{A}_{\rm A15}$ ($\bar{A}_{\rm A15}$ should be the weighted average of $A_{\rm Z14}$  and $A_{\rm Z12}$ ), this results suggests that the stability of different supperlattices does not simply follow the minimum-area rule as in foams~\cite{ziherl2001}.

\subsection{Distribution  of monomers around nanoparticles}
\begin{figure}[b!]
\begin{center}
	\includegraphics[width=0.45\textwidth]{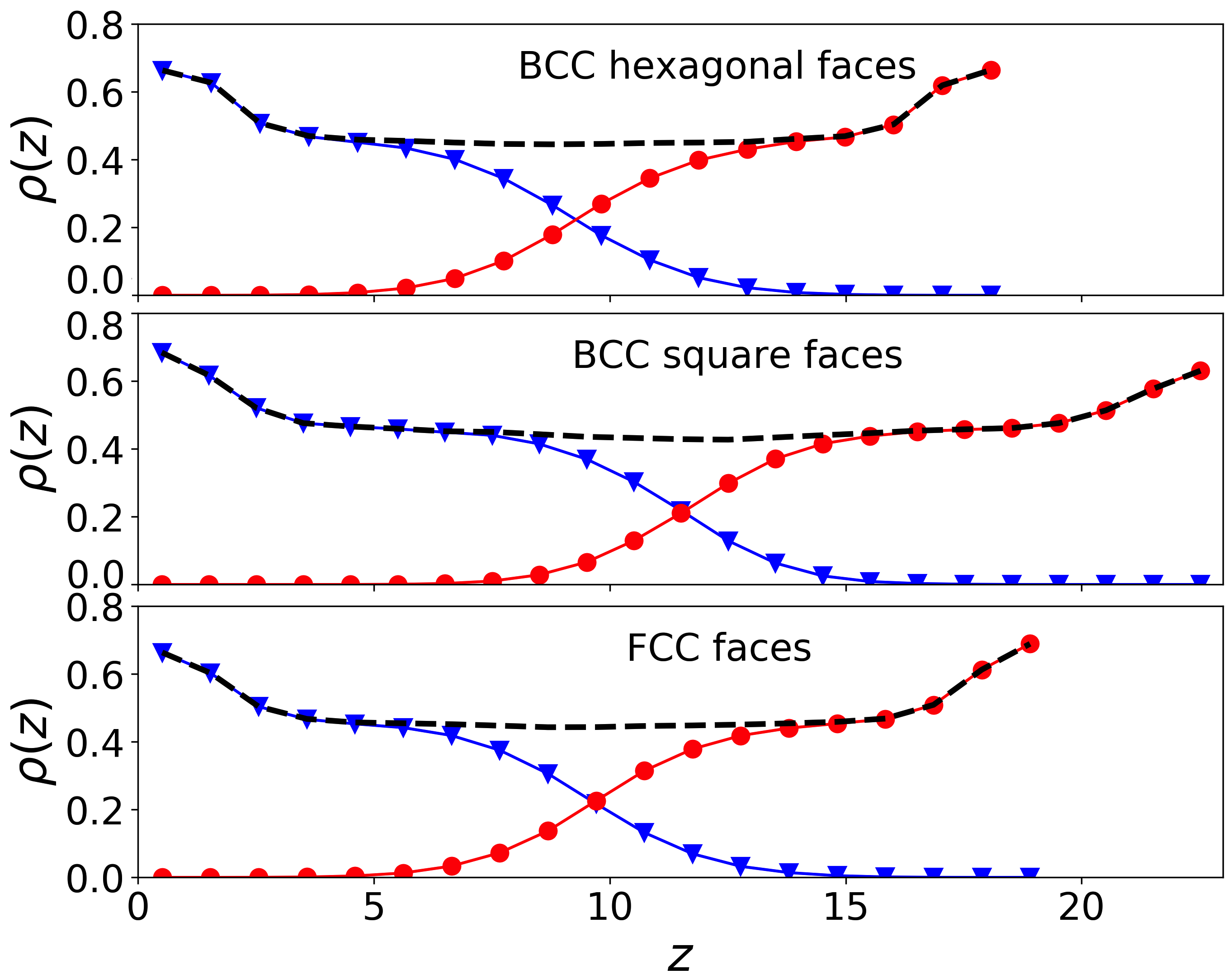}
	\captionof{figure}{The monomer density profile $\rho(z)$ along a crystallographic direction connecting the centers of two neighboring PGNPs, where $z$ is the distance from the surface of the central nanoparticle.  In BCC, two nonequivalent directions crossing either the hexagonal or the square faces of the truncated octahedron are considered, while  in FCC all the faces of rhombic dodecahedron are equivalent.   Contributions from the central nanoparticle (triangles) and the neighboring nanoparticle (circles) are plotted, together with their sum (dashed line).  $\rho(z)$ is obtained by sampling within a cylinder of radius $2\sigma$. $V=18000$ and $N_p = 40$. }
	\label{fig:rhoz}
\end{center}
\end{figure}

To reveal the underlying mechanisms for superlattice stability, we analyze the  spatial distribution of polymer chains around nanoparticle cores.  We first calculate monomer density profile $\rho(z)$ along a certain crystallographic direction perpendicular to polyhedral faces. In either BCC or FCC, the inter-particle space is filled with monomers from the two neighboring PGNPs structured in three layers (Fig.~\ref{fig:rhoz}). Close to the core,   monomer density $\rho(z)$ is higher than the bulk density $\bar{\rho}$ and decays from the maximum value corresponding to the high surface grafting density $\Sigma = 0.637  \sigma^{-2}$. At intermediate distances, there is a ``dry layer'' with $\rho(z) =\bar{\rho}$ and   monomers come from only one PGNP. Further away, coronas from two PGNPs start to overlap,  forming an ``interpenetration layer''~\cite{midya2020}. 

By symmetry, each shared polyhedral face bisects the center-to-center line at the location where monomer density from one PGNP equals to half of the total monomer density. This location can  also be read from half of the lattice constant $a$ along each  crystallographic direction (see Supporting Information for the value of $a$). Due to the interpenetration of chains, the faces of PGNP polyhedra   on supperlattice are strongly roughened, and the shape of PGNPs is more smoothened and spherical than  a perfect Wigner-Seitz polyhedron, as will be revealed below.

\subsection{Distribution  of end-to-end vector of grafted chains}
\begin{figure*}
\begin{center}
	\includegraphics[width=0.45\textwidth]{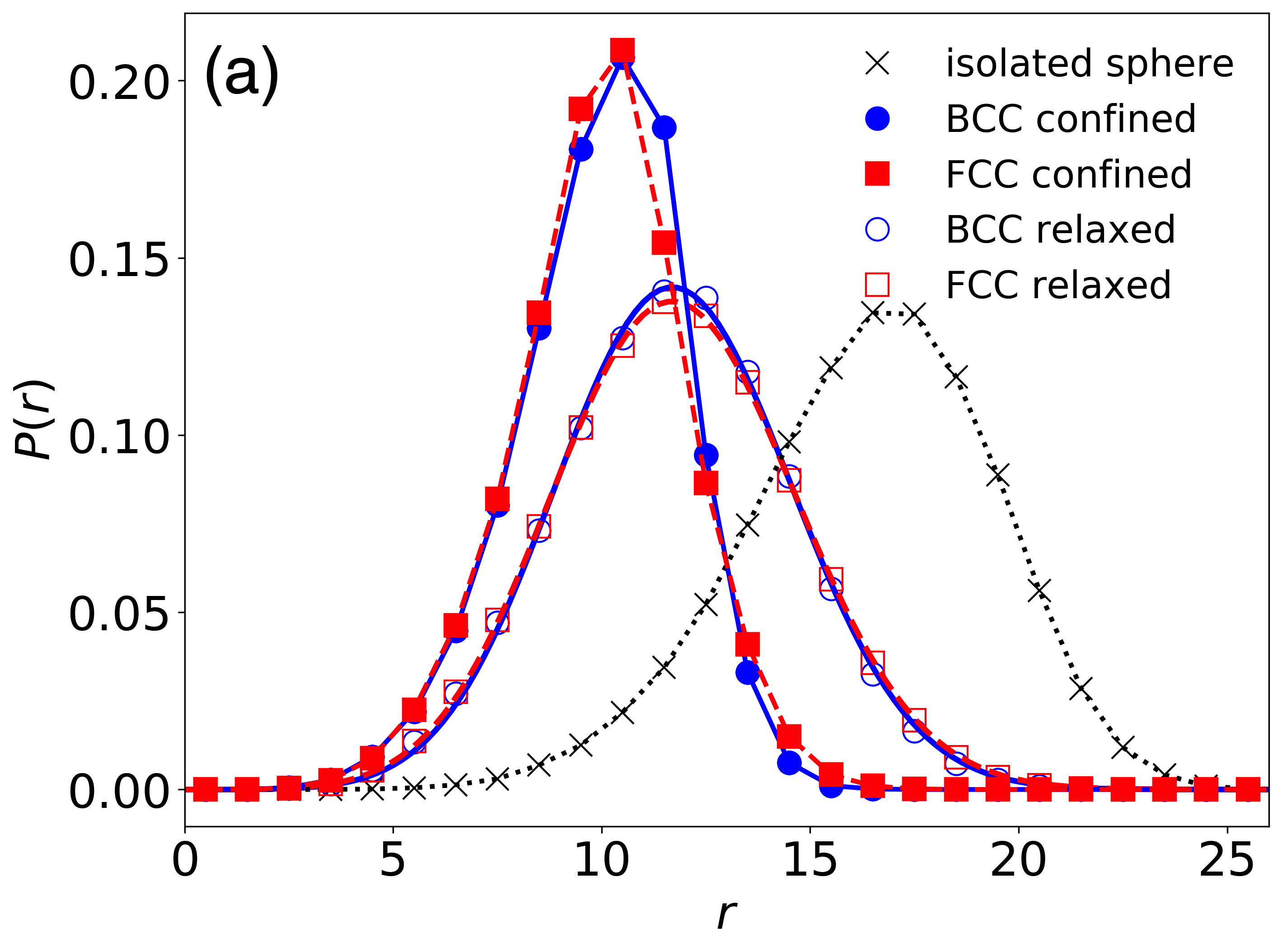}
		\includegraphics[width=0.45\textwidth]{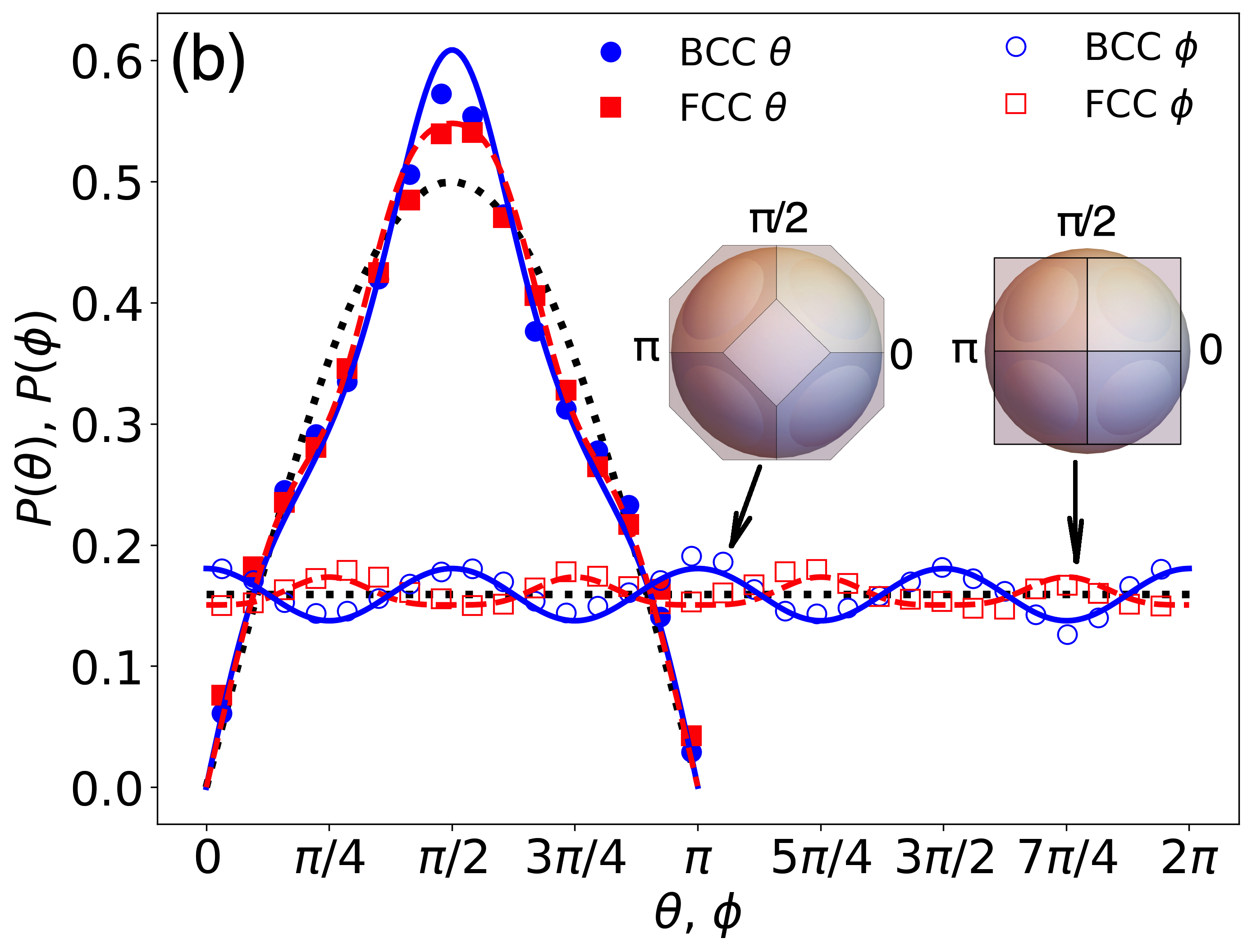}
	\captionof{figure}{Distribution of the  end-to-end vector ${\bf r} = (r, \theta, \phi)$ of all grafted polymers pointing from the anchoring monomer  to the  end monomer for BCC and FCC crystals at $V=18000$ and $N_p = 40$. (a) Radial distribution $P (r)$ of the length $r = |{\bf r} |$ in isolated spherical PGNPs (black dotted line), in compressed polyhedra after step I of thermodynamic integration (filled symbols) and in relaxed polyhedrons in crystal superlattice (empty symbols).  (c) Distribution of the  polar angle $\theta$ (filled symbols) and the azimuthal angle $\phi$  (empty symbols) of ${\bf r}$  compared with the case of a perfect spherical distribution (black dotted lines). Insets show the projection of truncated octahedron or rhombic dodecahedron onto the $xy$ plane, overlaid with the projection of a sphere of the same volume. }
	\label{fig:Rend}
\end{center}
\end{figure*}

We can quantify the states of grafted polymer chains by studying the distribution of their end-to-end vector ${\bf r}$, which points  from the first anchoring monomer to the last monomer on   each chain. The radial distribution $P(r)$ for  isolated spherical PGNPs, confined PGNPs under  wall compression (after step I of thermodynamic integration) and relaxed PGNPs of our interest are shown in Fig.~\ref{fig:Rend}a. Here  $P(r) d r$ is the probability for the chain end to fall in the spherical shell beween radius $r$ and $r+ dr$, regardless of chain orientation. For confined polyhedra (see inset in Fig.~\ref{fig:TIcurve}a),  $r$ are more narrowly distributed. FCC has a slightly higher $P(r)$ at  $r> 13$ and $r<10$, but BCC has more chains with $10< r < 13$. This trend qualitatively agrees with the solid angle distribution for perfect truncated octahedron and rhombic dodecahedron~\cite{ungar2003,goodfellow2015}. In the study of ligand-capped nanocrystals~\cite{goodfellow2015}, FCC Wigner-Seitz cell  is considered to have a more anisotropic  chain length (corona thickness) distribution, and to carry more entropy penalty than BCC when deformed from a hyperthetical sphere of the {\em same} volume $V$.  Although this argument about the entropy of chain length  distribution may explain the higher stability of our {\em confined} BCC structure with a relatively regular polyhedron cell,  it cannot resolve  the stability of different {\em relaxed}  structures as shown below. 

We observe that, after wall confinement is removed,  the distributions $P (r)$ of relaxed BCC and FCC are almost indistinguishable (empty symbols in Fig.~\ref{fig:Rend}a), which agrees with our expectation that interpenetration of chains blurs the boundary of Wigner-Seitz polyhedra. Because the radial distribution of chain length alone does not even distinguish the shape of the two polyhedra, we need to study the orientational distribution of ${\bf r} = (r,\theta,\phi)$, i.e.  $P(\theta)$ of   polar angle $\theta$  and $P(\phi)$ of azimuthal angle $\phi$.  In Fig.~\ref{fig:Rend}b, we plot $P(\theta)$ and $P(\phi)$ for relaxed polyhedra of BCC and FCC. Surprisingly,   both distributions have a larger deviation from the spherical case (black dotted lines) in BCC than in FCC.   The distribution  $P(\phi)$ suggests that chains are  more concentrated around the six square faces in truncated octahedron ($\phi =0, \pi/2, \pi, 3\pi/2$), which separate the central PGNP from its second nearest neighbors. In these directions of BCC, the lattice constant  is $2/\sqrt{3} = 1.155$ times of the nearest neighbor distance.  In the case of rhombic dodecahedron, chains are slightly more clustered around $\phi =\pi/4, 3\pi/4, 5\pi/4, 7\pi/4$, which correspond to the corner formed by four rhombus faces at a distance $\sqrt{2} = 1.414$ times of the nearest neighbor distance.
 
 \subsection{Entropy of end-to-end vector distribution}
After obtaining the radial $P (r)$ and angular  $P(\theta)$, $P(\phi)$ distribution of the end-to-end vector ${\bf r} = (r, \theta, \phi)$ of grafted chains, we can compute the (Gibbs) entropy $S$ of the overall distribution $P({\bf r})$ for a given (relaxed) Wigner-Seitz polyhedron. If  the total distribution of vector ${\bf r}$ can be factored as $P({\bf r}) = R(r) \Theta(\theta) \Phi(\phi)$, the corresponding normalization condition should be
 \begin{align*}
  \begin{aligned}
 1 &= \int d {\bf r} P({\bf r})\\
 &  = \int_0^\infty dr r^2  R(r) \int_0^\pi d\theta \sin \theta \Theta(\theta)  \int_0^{2\pi} d\phi \Phi(\phi).
 \end{aligned}
  \end{align*}
 By comparing with our three normalized distributions,  $\int\limits_0^\infty dr P(r)=1$,  $\int\limits_0^\pi d\theta  P(\theta) =1$ and $ \int\limits_0^{2\pi} d\phi P(\phi)$, we can relate $P(r) = r^2 R(r)$, $P(\theta) = \sin\theta \Theta(\theta)$ and $P(\phi) = \Phi(\phi)$ such that $P({\bf r}) = \frac{P(r)}{r^2} \frac{P(\theta)}{\sin\theta} P(\phi)  $.
 
 The entropy of the distribution $P({\bf r})$ can thus be expanded as
  \begin{align}
    \begin{aligned}
S&=  -k_B \int d {\bf r} P({\bf r})  \ln  P({\bf r}) \\
&=  -k_B  \int  dr d\theta d \phi P(r) P(\theta) P(\phi) \ln \left[ \frac{P(r)}{r^2} \frac{P(\theta)}{\sin\theta}  P(\phi)   \right] \\
& =S_r + S_\theta + S_\phi 
 \end{aligned}
 \end{align}
 where
   \begin{align}
    \begin{aligned}
S_r&=-k_B  \int_0^\infty dr   P(r)   \ln   \frac{P(r)}{r^2}   \\
S_\theta&=-k_B \int_0^\pi   d\theta   P(\theta)   \ln   \frac{P(\theta)}{\sin\theta}  \\
S_\phi & = -k_B  \int_0^{2\pi} d\phi P(\phi) \ln P(\phi).
 \end{aligned}
 \end{align}
 For spherically symmetric distributions,  $P(\theta) = \frac{1}{2} \sin\theta$ and $P(\phi) = \frac{1}{2\pi}$ (black dotted lines in Fig.~\ref{fig:Rend}b), which gives constant entropy terms $S_\theta({\rm sphere}) = k_B \ln 2$ and $S_\phi({\rm sphere}) = k_B \ln (2\pi)$. 
 
 We fit the sampled distributions of relaxed BCC and FCC polyhedra with following functions (the trivial phase shift of $\pi/4$  is applied to  the  $\phi$ angle of FCC before the fitting)
    \begin{align}
    \begin{aligned}
 P(r) &=     ar^2e^{-b(r-c)^2} \\
 P(\theta) & =     \frac{\sin\theta}{2} \left[c_1\cos(4\theta )+c_2\cos(8\theta)+ 1 \right] \\
 P(\phi) & = c_3\cos(4\phi)+c_4 \cos(8\phi)+\frac{1}{2\pi}.
 \end{aligned}
 \end{align}
 The fitting curves for relaxed BCC and FCC are shown in Fig.~\ref{fig:Rend}, from which the entropy of the distribution $P({\bf r})$ is calculated as in Table~\ref{table:S}.
 \begin{table}[h!]
 	\caption{Entropy $S$  of the distribution $P({\bf r})$ and its radial and angular contributions at $V = 18000$ and $V=34000$. Grafted polymers are flexible with FENE bonds.}
\begin{tabular}{lllll}
\hline 
$V$         & \multicolumn{2}{c}{18000} & \multicolumn{2}{c}{34000} \\
\hline 
            & BCC         & FCC         & BCC         & FCC         \\
\hline 
$S_r$      & 7.319       & 7.346       & 7.801       & 7.806       \\
$S_\theta$ & 1.833       & 1.837       & 1.830       & 1.837       \\
$S_\phi$   & 0.684       & 0.689       & 0.681       & 0.690       \\
\hline
$S$         & 9.833       & 9.879       & 10.300      & 10.319    \\
\hline
\end{tabular}
 \label{table:S}
\end{table}

It can be seen that the total entropy $S$ and its radial or angular contributions of BCC structure are all lower or close to those of FCC. This disagrees with the previous knowledge that the higher stability of BCC superlattice can be explained by the entropy of chain length alone~\cite{goodfellow2015}. It is worth emphasizing that chain interpenetration makes the radial distribution $P(r)$ of chain length almost identical for BCC and FCC polyhedra. We thus argue that the free energy and  stability of various superlattices should be dominated by energy and other entropy contributions, for instance, the configuration entropy of individual monomers.

 \subsection{Potential of mean force between two PGNPs}
 Although designed to calculate the free energy  of superlattices of  PGNPs, the current method can be straightforwardly modified to compute the potential of mean force (PMF) between a pair of   PGNPs. Only one confining wall per PGNP is needed at the bisecting plane of the center-to-center line and only the two interacting PGNPs need to be considered under open boundary conditions in step II of the thermodynamic integration. Previously, the PMF  has been estimated from the radial distribution function~\cite{striolo2007}, interaction energy~\cite{verso2011} and  force~\cite{meng2012} between two PGNPs. The similar problem has been  studied in the context of polymer brush~\cite{milner1988,milner1991}  and   steric stabilization of colloidal particles~\cite{centeno2014}. 
  \begin{figure}
\begin{center}
	\includegraphics[width=0.4\textwidth]{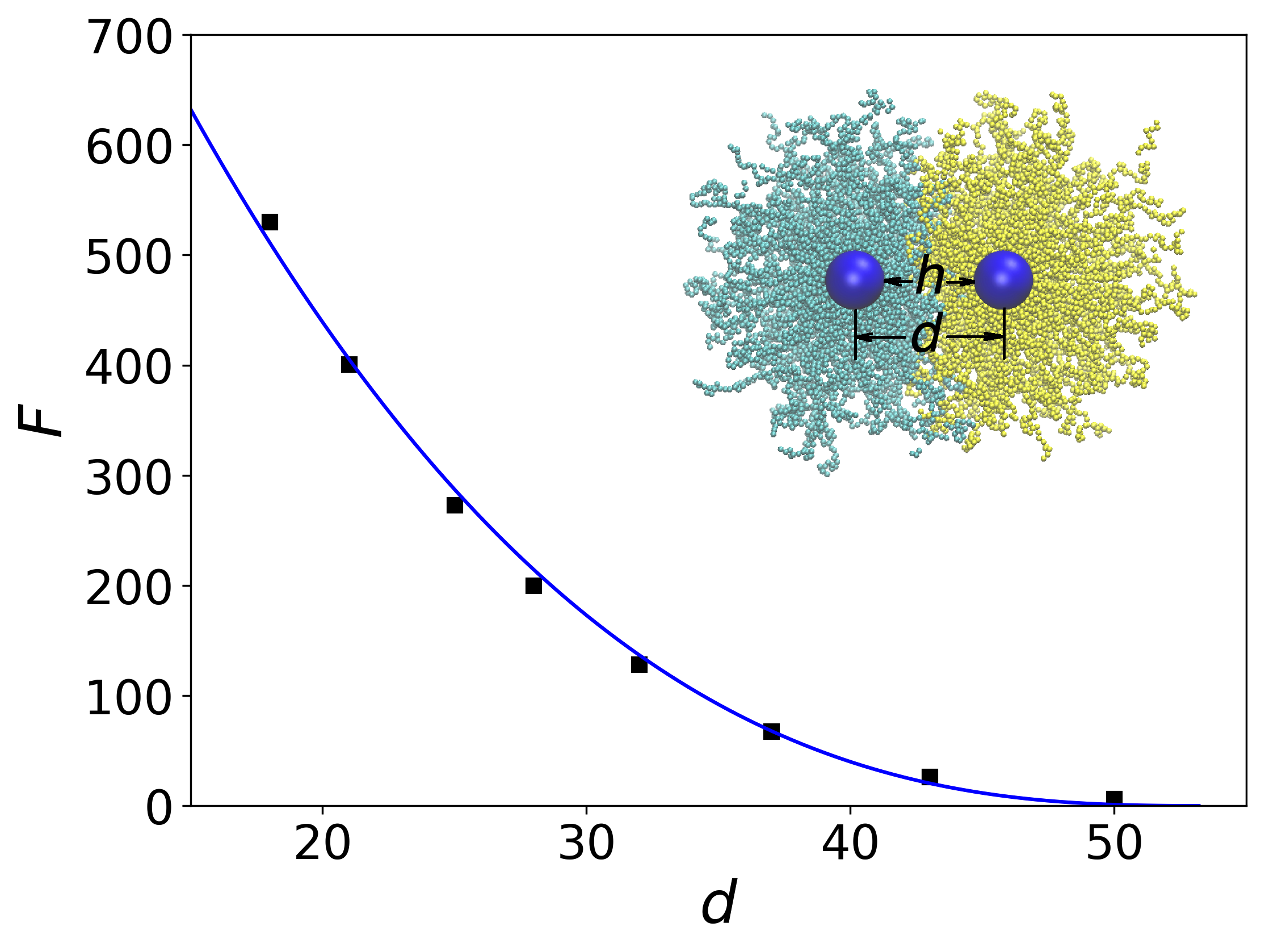}
	\captionof{figure}{Potential of mean force $F$ between a pair of PGNPs with $N_p = 60$ as a function of the center-to-center distance $d$, fitted by a Hertzian-like  potential law $F \propto  (1- \frac{d}{2 R_0})^{\alpha}$ with $\alpha = 2.6$. The radius of isolated spherical PGNPs is $R_0 = 26.6$. }
	\label{fig:PMF}
\end{center}
\end{figure}

We test this idea on a system with chain length $N_p = 60$ by calculating the PMF $F$  as a function of the center-to-center distance $d$. Obviously,  the two PGNPs only start to experience  a repulsion when $d < 2R_0$, where $R_0$ is the radius of the isolated spherical PGNP with that chain length. Instead of the combined power law and inverse power law in terms of surface-to-surface distance $h$ (or brush height $h/2$)~\cite{milner1988,centeno2014}, our results show that  the PMF can follow  a simpler Hertzian-like  repulsion   $F \sim  (1- \frac{d}{2 R_0})^{\alpha}$ with a fitting exponent $\alpha = 2.6$~\cite{johnson1987}.

\section{Conclusion}
 In this work, we present a simulation method that computes the free energy cost to deform a spherical PGNP into a Wigner-Seitz polyhedron followed by relaxation in a superlattice with certain crystalline symmetry. By applying this thermodynamic integration scheme on various systems, we confirm that BCC is   more stable that FCC and A15 under most simulation conditions, but only by a small  free energy difference.   Comparison of polyhedral surface area  and entropy of corona chain distribution suggests that none of these factors can explain the  higher stability of BCC superlattice alone. The total free energy must be determined by the intricate interplay between energy and entropy under the superlattice geometry.

The current method can further be adapted to study the self-assembly of PGNPs in polymer melts~\cite{koh2020} or other deformable mesoparticles such as block copolymer micelles. As in the last example of PMF calculation, it would also be interesting to study polymer brush interactions or free energy cost to insert a nanoparticle into polymer brush~\cite{milchev2008} with our method.

%

\balance
 
\bibliography{pgnp}  

\end{document}